\documentclass{article}
\usepackage{graphicx}
\newcommand{\beq}{\begin{equation}}
\newcommand{\eeq}{\end{equation}}
\newcommand{\beqa}{\begin{eqnarray}}
\newcommand{\eeqa}{\end{eqnarray}}
\newcommand{\lslash}[1]{#1\llap/}

\newcommand{\Tr}{\mathop{\rm Tr}}

\renewcommand{\Re}{\mbox{Re}}

\newcommand{\Eq}[1]{Eq.\ (\ref{#1})}
\newcommand{\Eqs}[2]{Eqs.\ (\ref{#1}) and (\ref{#2})}

\newcommand{\Ref}[1]{Ref.\ \cite{#1}}
\newcommand{\Fig}[1]{Fig.\ \ref{#1}}

\newcommand{\Sec}[1]{Section\ \ref{#1}}

\newcommand{\pieff}{\hat\pi}
\newcommand{\para}{{\scriptscriptstyle\parallel}}

\title{Thermal Field Theory in a wire: Applications of
Thermal Field Theory methods to the propagation
of photons in a one-dimensional plasma}

\author{Jos\'e F. Nieves \\
Laboratory of Theoretical Physics\\
Department of Physics, P.O. Box 70377\\
University of Puerto Rico\\
San Juan, Puerto Rico 00936-8377}

\begin{document}
\maketitle

The Thermal Field Theory methods are applied to calculate
the dispersion relation of the photon propagating modes in a strictly
one-dimensional ideal plasma. The electrons are treated as a gas of particles
that are confined to a one-dimensional tube or wire,
but are otherwise free to move, without reference to the
electronic wave functions in the coordinates that are transverse to the
idealized wire, or relying on any features of the electronic structure.
The relevant photon dynamical variable is an effective field in which
the two space coordinates that are transverse to the wire are collapsed.
The appropriate expression for the photon \emph{free-field} propagator
in such a medium is obtained, the one-loop photon self-energy
is calculated and the (longitudinal) dispersion relations are determined and
studied in some detail. Analytic formulas for the dispersion relations
are given for the case of a degenerate electron gas, and the results
differ from the long-wavelength formula that is quoted in the literature
for the strictly one-dimensional plasma. The dispersion relations
obtained resemble the linear form that is expected in realistic
quasi-1D plasma systems for the entire range of the momentum,
and which have been observed in this kind of system in recent experiments.
%
%
\section{Introduction}
\label{sec:introduction}

The application of thermal field theory (TFT)\cite{weldon:cov,niemi1,niemi2}
to study lower dimensional systems is of interest for several reasons.
Some of them have to do with the intrinsic interest and potential applications
that they may have in condensed-matter and other related branches of physics.
On the other hand, there has recently been suggestions
that lower dimensional systems may be relevant at a fundamental
level. One idea is that the effective dimensionality of the
space we live in depends on the length scale being probed and
in particular at short scales the space is lower
dimensional\cite{stojkovic1,stojkovic2,stojkovic3}, and lower dimensional
systems may also emerge as some extension of
the standard model of particle physics\cite{arnold}.
In such contexts, the TFT formulation for
lower dimensional systems or models
can have phenomenological applications for particle physics and cosmology
as well.

Here we use the TFT methods to study the electromagnetic properties
of a one-dimensional plasma, in which the electrons are confined
to a line (i.e., an ideal one-dimensional wire which can be taken
to be the $z$ axis). These systems have been studied in recent
experiments\cite{kukushkin}, and earlier from a theroretical
point of view\cite{sarma,bonitz}. The analogous non-abelien systems
have also been considered\cite{zonca}.
The quantities of interest are the dispersion relations and damping
of the propagating photon modes and,
from the TFT point of view, the quantities to be determined are the
free-field propagator for the photon effective field in the medium
and the self-energy, from which the dispersion relations can be
obtained.

The same method has been applied to the 2D plane sheet in \Ref{nieves:layer}.
As emphasized there, this is not the same thing as what
is usually called $QED_3$ (or $QED$ in 2+1 dimensions), which had
been studied previously in the
literature\cite{KleinKreisler:1994wr,Hott:1998qu}.
In $QED_3$, the system, with regard to the space coordinates,
has cylindrical symmetry, and the physics, being independent of the
$z$ coordinate, can be studied by considering a two-dimensional cross section.
An important consequence of this difference
is that the propagation of the photon in the layer
is described by an \emph{effective} field which has a corresponding
\emph{free-field} propagator
that is very different from the usual one. The photon propagator is an
important quantity because its inverse determines the bilinear part
of the effective action or, equivalently, the equation of motion for the
effective field, from which the dispersion relations and wave
functions of the propagating modes can be obtained.

We follow here a similar procedure for the wire. As we will see,
the photon free-field propagator in the present case is
ultraviolet logarithmically divergent. The existence of this kind of divergence
in the strictly 1D plasma has been known for some time\cite{sarma}.
The workaround has been to introduce a cut-off parameter
and evaluate the divergent integrals in terms of it. However, this
procedure leads to a long-wavelength dispersion relation that
is valid only for a very small range of the momentum $\kappa$.
The alternative has been to consider more realistic models
using the fact that a real quasi-1D plasma has
some finite radial width, and then take into account the
electronic bound state wave functions in the directions perpendicular
to the wire. Since the calculations of this type
are typically numerical ones, and they involve some knowledge or modeling
of the electronic structure of the system under consideration,
they are not applicable to more general situations such as,
for example, the relativistic and high energy limits.
Thus, there is no treatment of the strictly 1D ideal plasma system
that models the results predicted by studying realistic quasi-1D
plasma systems.

This work fills this gap. Here we insist in considering the strictly
1D ideal system of free electrons. We determine the appropriate expression
for the photon free-field propagator and, using the TFT rules,
we obtain the 1-loop formula for the photon self-energy in the medium.
The formula is given, as usual as an integral over the electron momentum
distribution function. The general expression for the dispersion relation
is obtained and explicit formulas are given by considering specifically
the case of a degenerate electron gas. The result differs
from the cut-off dependent long-wavelength formula that is quoted in the
literature for the strictly 1D system. The dispersion relation obtained
resembles the linear form that has been observed in the recent
experiments in systems of this type\cite{kukushkin} and which is
expected on the basis of the numerical calculations that take into account
the electronic structure and finite-width of a real quasi-1D
system. The result obtained here for the degenerate, strictly 1D system,
has a leading term $v_F \kappa$, where $v_F$ is the Fermi velocity
of the electrons, with logarithmic corrections proportional
to $e^2 v^2_F/2\pi^2$.

The logarithmic ultraviolet divergence
of the free-field photon propagator is treated by noticing that its
derivative is finite and calculable.
The propagator, and the physical quantities, then depend on an
unknown mass scale parameter $\mu$ that appears as an integration constant.
Our point of view is that this leads to an effective theory,
and the results obtained for the physical quantities of interest
are applicable to a wide range of real quasi-1D systems,
in the situations in which they can be idealized as a
1D plasma of free electrons, independently of the particular electronic 
structure, width or geometry in the perpendicular directions of such systems.
Those details are parametrized by the single parameter 
$\mu$. While $\mu$ cannot be determined, this strategy is
very economical since it is the only unknown free parameter of the model.

This paper is organized as follows. In \Sec{sec:kinematics} we summarize
the notation for the kinematic variables used throughout. In \Sec{sec:model}
the assumptions that define the model are stated precisely. In particular,
the photon effective field that will be treated as the relevant dynamical
variable is identified, along with its interaction with the electron field.
In \Sec{sec:photonpropagation} the photon free field propagator is determined
and the equation for the dispersion relation of the propagating modes
is written in terms of its inverse and the photon self-energy.
In \Sec{sec:disprel} the one-loop calculation of the photon self-energy
is carried out and the dispersion relations are studied in some detail
by considering specifically the case of a degenerate electron gas.
%
%
\section{Notation and kinematics}
\label{sec:kinematics}

We denote by $u^\mu$ the velocity four-vector of the medium.
Adopting the frame in which the medium is at rest, we set
\beq
u^\mu = (1, \vec 0) \,,
\eeq
and from now on all the vectors refer to that frame.
We introduce the four-vector
\beq
n^\mu = (0, \vec n) \,,
\eeq
where $\vec n$ is the unit vector along the wire,
and denote the momentum four-vector of a photon that propagates in the wire
by
\beq
\label{kparadef}
k^\mu_\para = (\omega, \vec \kappa_\para) \,,
\eeq
where
\beq
\label{veckappa}
\vec \kappa_\para = \kappa \vec n\,,
\eeq
and its square
\beq
\label{kparasquare}
k^2_\para = \omega^2 - \kappa^2 \,.
\eeq
Since neither $u^\mu$ nor $n^\mu$ is orthogonal to $k^\mu_{\para}$,
it is useful to define  the combination
\beq
\tilde u_{\mu} \equiv u_\mu - \frac{\omega k_{\para\mu}}{k^2_\para}\,,
\eeq
which satisfies
\beq
k_\para\cdot\tilde u = 0\,.
\eeq

Any given four dimensional vector $a^\mu = (a^0, \vec a)$
can be decomposed in the form
\beq
\label{adecomp}
a^\mu = a^\mu_\para + a^\mu_\perp \,,
\eeq
where
\beq
a^\mu_\para = (a^0, \vec a_\para)\,,\qquad
a^\mu_\perp = (0, \vec a_\perp)\,,
\eeq
with
\beq
\vec a_\para = (\vec a\cdot \vec n)\vec n\,,\qquad
\vec a_\perp = \vec a - \vec a_\para\,.
\eeq
Without loss of generality, we can take the $z$ axis to point
along $\vec n$, and with that convention the components of
$\vec a_\para$ and $\vec a_\perp$ are
\beqa
a^i_\para & = & (0, 0, a^3)\,,\nonumber\\
a^i_\perp & = & (a^1, a^2, 0) \,.
\eeqa

It is useful to introduce the tensors 
\beqa
\label{RQ}
Q_{\mu\nu} & = & \frac{\tilde u_\mu \tilde u_\nu}{\tilde u^2}\,,\nonumber\\
R_{\mu\nu} & = & g_{\mu\nu} - \frac{k_{\para\mu} k_{\para\nu}}{k^2_\para} - 
Q_{\mu\nu}\,,
\eeqa
which are transverse to $k^\mu_\para$, and define
\beq
\label{gparallel}
g_{\para\mu\nu} = g_{\mu\nu} - R_{\mu\nu}\,.
\eeq
Noticing that $R$ satisfies
\beq
u^\mu R_{\mu\nu} = n^\mu R_{\mu\nu} = 0\,,
\eeq
the decomposition in \Eq{adecomp} can be accomplished by writing
\beqa
a_{\perp\mu} & = & R_{\mu\nu} a^\nu\,,\nonumber\\
a_{\para\mu} & = &  g_{\para\mu\nu} a^\nu\,.
\eeqa
Moreover, if $a^\mu$ is transverse to $k^\mu_\para$, that is
\beq
a\cdot k_\para = 0\,,
\eeq
we then have
\beq
a_{\para\mu} = Q_{\mu\nu} a^\nu \,,
\eeq
and in particular
\beq
\label{propto}
a_{\para\mu} \propto \tilde u_\mu \,.
\eeq
%
%
%
\section{The Model}
\label{sec:model}

We wish to emphasize that these two systems, to which we refer
as the \emph{layer} and the \emph{wire}, are not the same thing as what
are usually called $QED_3$ (or $QED$ in 2+1 dimensions) and 
$QED$ in 1+1 dimensions, respectively. For example in $QED_3$,
which has been studied previously in the
literature\cite{KleinKreisler:1994wr,Hott:1998qu},
the system has cylindrical symmetry in the spatial coordinates,
and therefore the physics, being independent
of the $z$ coordinate, can be studied by considering a two-dimensional
cross section. Thus, for example, the \emph{electron} in $QED_3$ is really
a line of charge in the three-dimensional world, and the Coulomb potential
between two such \emph{electrons} is logarithmic. In contrast, in
the system that we considered, the electron is an ordinary point charge,
which is confined to the $z = 0$ plane, but the Coulomb potential between
two electrons is the usual $1/r$ potential.

\subsection{The electron field}

We envisage the system as a tube along the $z$ axis of length $L$ and 
cross sectional area $L^2_\perp$ in which the electrons are confined
but otherwise free to move, and eventually the limits
$L \rightarrow \infty$ and $L_\perp \rightarrow 0$ are taken.
In the Furry picture, the one-particle electron wavefunction is a product of
a plane wave in the $z$ direction times some function $H(\vec x_\perp)$ 
in the perpendicular directions. While in principle the function $H$ is
characterized by some quantum number, the assumption
is that in the $L_\perp \rightarrow 0$ limit only the lowest state survives. 
Specifically, the model is based on the assumption that the electron
wavefunctions are such that, when that limit is taken,
the electron current density operator reduces to
\beq
\label{jdef}
j^\mu(x) = \delta^{(2)}(\vec x_\perp) \overline{\hat\psi}(x_\para)
\gamma^\mu_\para\hat\psi(x_\para) \,,
\eeq
where, in the free field case,
\beq
\label{psipara}
\hat\psi(x_\para) = \int\frac{dp}{(2\pi) 2E}\left[
a(\vec p_\para,s) u(\vec p_\para,s) e^{-ip_\para\cdot x_\para}
+
b^\ast(\vec p_\para,s) v(\vec p_\para,s) e^{ip_\para\cdot x_\para}\right]\,,
\eeq
with
\beqa
\label{pparadef}
p^\mu_\para & = & (E,\vec p_\para)\,,\nonumber\\
\vec p_\para & = & p\vec n \,,\nonumber\\
E & = & \sqrt{p^2 + m^2}\,.
\eeqa
The spinors $u$ are the standard Dirac spinors normalized such that
\beq
u\bar u = 2m\,,
\eeq
and the creation and annihilation operators satisfy
\beq
\left\{a(\vec p_\para,s),a^\ast(\vec p^{\,\prime}_\para,s^\prime)\right\} = 
(2\pi) 2E\, \delta(p - p^\prime)
\delta_{s,s^\prime} \,,
\eeq
with analogous relations for the spinors $v$ and the $b$ operators.

The thermal propagators for the field $\hat\psi$ are given by the
familiar formulas\cite{weldon:cov,niemi1,niemi2},
\beqa
\label{fermionpropagator}
S_{11}(p_\para) & = & (\lslash{p}_\para + m_e)\left[
\frac{1}{p^2_\para - m^2_e + i\epsilon}
+ 2\pi i\delta(p^2_\para - m^2_e)\eta_e(p_\para)\right] \,,\nonumber\\
S_{22}(p_\para) & = & (\lslash{p}_\para + m_e)\left[
\frac{-1}{p^2_\para - m^2_e - i\epsilon}
+ 2\pi i\delta(p^2_\para - m^2_e)\eta_e(p_\para)\right] \,,\nonumber\\
S_{12}(p_\para) & = & (\lslash{p}_\para + m_e)2\pi i\left[
\eta_e(p_\para) - \theta(-p_\para\cdot u)\right]\,,\nonumber\\
S_{21}(p_\para) & = & (\lslash{p}_\para + m_e)2\pi i\left[
\eta_e(p_\para) - \theta(p_\para\cdot u)\right]\,,
\eeqa
where
\beq 
\label{etae} 
\eta_e(p) = \theta(p\cdot u)f_e(p\cdot u) +
\theta(-p\cdot u)f_{\bar e}(-p\cdot u)\,, 
\eeq
with
\beqa 
\label{fe}
f_e(x) & = & \frac{1}{e^{\beta(x - \mu_e)} + 1} \nonumber\\
f_{\bar e}(x) & = & \frac{1}{e^{\beta_e(x + \mu_e)} + 1}
\eeqa
and $\theta(x)$ is the step function.
Here $\beta_e$ and $\mu_e$ are the inverse temperature and the
chemical potential of the electron gas, respectively.

The total number of particles in the gas can be calculated from
\beqa
N & = & \int d^3x\,j^0(x) \nonumber\\
& = & L \int\frac{dp}{(2\pi)}
\mbox{Tr}\left[S_{11}(p_\para)\gamma^0\right]\,,
\eeqa
where we have used \Eq{jdef} and we have set $\int dz \rightarrow L$.
Using formulas given above for the propagator, this yields
\beq
N/L = n_e + n_{\bar e} \,,
\eeq
where
\beq
\label{ne}
n_{e,\bar e} = 2\int\frac{dp}{(2\pi)}
\frac{1}{e^{\beta(E \mp \mu_e)} + 1}
\eeq
represent the linear density of electrons and positrons, respectively.

\subsection{The photon effective field}
\label{sec:photoneffectivefield}

Using \Eq{jdef} the usual interaction Lagrangian term
$j\cdot A$ yields the following term in the action
\beq
\label{Sint}
S_{\mbox{int}} = - e\int d^2x_\para\;\overline{\hat\psi}\gamma^\mu_\para
\hat\psi \hat A_\mu \,,
\eeq
where
\beq
\hat A_{\mu} \equiv \left. A_{\para\mu} \right|_{\vec x_\perp = 0} \,.
\eeq
This indicates in particular that the \emph{transverse} component 
$A^\mu_\perp$ decouples.  Thus, regarding $\hat A_\mu$ as the effective field
for the photon, our goal is to determine its effective action,
or equivalently its equation of motion, including the thermal corrections.
Formally, this involves integrating out all the dynamical field
variables except $\hat A_\mu$ itself. 

Adapting the functional method of quantization of the electromagnetic 
field\cite{peskin} to the present model, 
a convenient way to proceed is to introduce in the action an 
external current of the form
\beq
\label{Jdef}
J^\mu(x) = \delta^{(2)}(\vec x_\perp) J^\mu_\para(x_\para) \,,
\eeq
with $J^\mu_\para(x_\para)$ satisfying
\beq
\label{Jtranscond}
\partial_\para\cdot J_\para(x_\para) = 0\,,
\eeq
where $\partial^\mu_\para = (\frac{\partial}{\partial x_0},
-\frac{\partial}{\partial\vec x_\para})$.
\Eq{Jtranscond}, which in turns implies,
\beq
\partial\cdot J = 0 \,.
\eeq
ensures that the source term is selecting the gauge invariant
(i.e., transverse to the photon momentum four-vector) part of the
\emph{longitudinal} component $A^\mu_\para$.

The \emph{classical} field $A^{(J)}_\mu$,
in the presence of both the external current $J_{\para\mu}$ and the interaction
given by $S_{\mbox{int}}$ in \Eq{Sint}, is then defined by
\beq
A^{(J)}_\mu = \frac{1}{Z}\frac{i\delta Z}{\delta J^\mu_\para} \,,
\eeq
where $Z$ is the generating functional.
%
%
\section{Photon propagation in the wire}
\label{sec:photonpropagation}

\subsection{Photon free-field propagator}
\label{sec:photonpropagator}

The question here is, what is the propagator associated with the
photon effective field $\hat A_{\mu}$?
Following the usual argument, the generating functional for the
photon free field in the wire is
\beq
Z \propto \exp\left\{-\frac{i}{2}\int d^2x_\para d^2x^\prime_\para
J^\mu_\para(x_\para)\hat\Delta_{F\mu\nu}(x_\para - x^\prime_\para)
J^\nu_\para(x^\prime_\para)\right\}\,,
\eeq
where $\hat\Delta_{F\mu\nu}(x_\para - x^\prime_\para)$ is obtained from the
standard photon propagator $\Delta_{F\mu\nu}(x - x^\prime)$
by setting the coordinates normal to the wire
($\vec x_\perp$ and $\vec x_\perp^\prime$) equal to zero,
as implied by the delta function in \Eq{Jdef}. Therefore,
taking into account \Eq{Jtranscond} and remembering \Eq{propto},
the free-field propagator in the wire is given, in momentum space, by
\beq
\hat\Delta_{F\mu\nu}(k_\para) = Q_{\mu\alpha}
Q_{\nu\beta}\left(
\int\frac{d^2k_\perp}{(2\pi)^2}\; \Delta_F^{\alpha\beta}(k)\right) \,,
\eeq
where $Q_{\mu\nu}$ has been defined in \Eq{RQ} and,
in the integrand, the momentum vector $k$ is decomposed in the form
\beq
k_\mu = k_{\perp\mu} + k_\para\,,
\eeq
with $k_{\para\mu}$ as given in \Eq{kparadef} and
$k^\mu_\perp = (0, \vec \kappa_\perp)$. Writing
\beq
\Delta_{F\mu\nu}(k) = \frac{-g_{\mu\nu}}{k^2 + i\epsilon} + 
\mbox{gauge-dependent terms} \,,
\eeq
we then obtain
\beq
\label{effectivefreeprop}
\hat \Delta_{F\mu\nu}(k_\para) = -\hat\Delta(k_\para) Q_{\mu\nu}\,,
\eeq
where
\beq
\hat\Delta(k_\para) = \int\frac{d^2\kappa_\perp}{(2\pi)^2}
\frac{1}{k^2_\para - \kappa^2_\perp + i\epsilon} \,.
\eeq
This integral is ultraviolet logarithmically divergent and whence it cannot
be computed using the above formula literally without modification. This is
of course due to our insistence on considering the strictly infinitely thin
wire. However, rather than give up and resort to the fact that a realistic
wire has some finite width and incorporating the electronic 
wave functions would soften the integral, we proceed as follows. 

We notice that its derivative
\beq
\frac{\partial\hat\Delta(k_\para)}{\partial k^2_\para} = 
-\int\frac{d^2\kappa_\perp}{(2\pi)^2}
\frac{1}{\left(k^2_\para - \kappa^2_\perp + i\epsilon\right)^2} \,,
\eeq
is finite and therefore, by a straightforward evaluation, 
\beq
\frac{\partial\hat\Delta(k_\para)}{\partial k^2_\para} = 
\frac{1}{4\pi}\frac{1}{k^2_\para + i\epsilon} \,.
\eeq
Thus, by direct integration this implies that
\beq
\label{regDelta}
\hat\Delta(k_\para) = \frac{1}{4\pi}\log\left(\frac{k^2_\para + i\epsilon}
{\mu^2}\right) \,,
\eeq
where $\mu^2$, which appears is a constant of integration,
is a parameter that we cannot determine further.

The existence of the divergences of the strictly 1D plasma has been
known in the plasma physics literature for some time. The workaround has been
to resort to the fact that a real quasi-1D plasma has
some finite (radial) width, and then take into account the
electronic bound state wave functions in the directions perpendicular
to the wire\cite{sarma}. Thus, since the calculations of this type
have been numerical ones, and they involve some knowledge or modeling of the
electronic structure of the system under consideration,
they are not applicable to more general situations such as,
for example, the relativistic and high energy limits.

Our point of view is that this approach leads us to an effective theory
that is valid for energy scales $(\omega, \kappa, \mu)$
less than some mass scale $\Lambda$ which is of the order
of the inverse length of a radial dimension of the wire.
The results obtained for the physical
quantities of interest using the effective theory
can then be applicable to a wide range of real quasi-1D systems,
in the situations in which they can be idealized as a
1D plasma of free electrons, independently of the particular electronic 
structure, width or geometry in the perpendicular directions of such systems.
All those details are hidden and parametrized by the single parameter 
$\mu$, which we cannot determine, reflecting our ignorance of those details.
On the other hand, this strategy is very economical when we take
into account the fact that $\mu$ is the only unknown parameter of the model.

\subsection{Photon self-energy and equation of motion}
\label{sec:eqofmotion}

Denoting the photon self-energy in the medium by $\pieff_{\mu\nu}$,
the bilinear part of the effective action for $A^{(J)}_\mu$
is then given, in momentum space, by
\beqa
S^{(2)} & = & \int \frac{d^3 k_\para}{(2\pi)^3}\left\{
\frac{1}{2}A^{(J)\ast}_\mu(k_\para) \left[D^{\mu\nu}(k_\para) +
\hat\pi^{\mu\nu}(k_\para)\right] A^{(J)}_\nu(k_\para)\right.\nonumber\\
&&\mbox{} \left. -A^{(J)\ast}(k_\para)\cdot \hat J(k_\para)\right\} \,,
\eeqa
where $D^{\mu\nu}(k_\para)$ is defined by
\beq
\hat \Delta_F^{\mu\lambda}(k_\para) D_{\lambda\nu}(k_\para) = 
Q^\mu_\nu \,.
\eeq
The equation of motion for the classical field in the absence of the external
current is then
\beq
\label{eqofmotion}
\left[D^{\mu\nu}(k_\para) +
\hat\pi^{\mu\nu}(k_\para)\right] A^{(0)}_\nu(k_\para) = 0\,.
\eeq
Since on one hand \Eq{effectivefreeprop} implies that
\beq
D_{\mu\nu}(k_\para) = -\hat\Delta^{-1}(k_\para) Q_{\mu\nu}\,,
\eeq
where $\hat\Delta(k_\para)$ is given in \Eq{regDelta}, while
in the other hand,
as we will verify, $\pieff_{\mu\nu}$ is of the form
\beq
\label{pieffform}
\pieff_{\mu\nu}(k_\para) = \pieff(k_\para) Q_{\mu\nu} \,,
\eeq
\Eq{eqofmotion} implies the condition
\beq
\label{dispreleq}
\hat\Delta^{-1}(k_\para) - \pieff(k_\para) = 0\,,
\eeq
which determines the dispersion relations of the propagating modes.

Furthermore, since in this work we are concerned only with 
the real part of the dispersion
relation, in order to determine $\pieff$ we need to calculate only 
the $11$ element of the thermal self-energy matrix $\pi^{(ab)}_{\mu\nu}$,
in terms of which
\beq
\Re\, \pieff_{\mu\nu}(k_\para) = \Re\, \pi^{(11)}_{\mu\nu}(k_\para)\,.
\eeq
Taking the real part in \Eq{regDelta}, the real part of the 
dispersion relation is then obtained from
\beq
\label{redispreleq}
4\pi\left(\log\left|\frac{k^2_\para}{\mu^2}\right|\right)^{-1}
- \Re\,\pieff = 0\,.
\eeq
%
%
%
\section{Self-energy and dispersion relations}

\subsection{One-loop formula for the self-energy}
\label{sec:oneloopselfenergy}

\begin{figure}
\begin{center}
\includegraphics[bb=198 578 400 640]{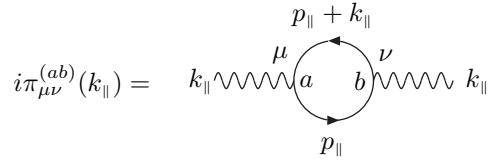}
\end{center}
\caption[]{One-loop diagram for the photon thermal self-energy matrix.
\label{fig:oneloopdiagram}
}
\end{figure}

Referring to Fig. \ref{fig:oneloopdiagram}, the one-loop formula for the
the $11$ element of the photon self-energy matrix is
\beq
\label{pi11}
i\pi^{(11)}_{\mu\nu}(k_\para) = e^2 \int\frac{d^2p_\para}{(2\pi)^3}
\Tr\gamma_{\para\mu} iS_{11}(p_\para + k_\para)
\gamma_{\para\nu} iS_{11}(p_\para) \,.
\eeq
When the formula for $S_{11}$ given in
\Eq{fermionpropagator} is substituted in \Eq{pi11}, there are
three types of terms. The term that contains two factors of $\eta_e$
contributes only to the imaginary part of the self-energy and,
since we restrict ourselves here to the real part, we do not consider
it further. The remaining terms then yield
\beq
\label{pimdef}
\mbox{Re}\,\pi^{(11)}_{\mu\nu} = \pi^{(0)}_{\mu\nu} + \pi^{(m)}_{\mu\nu}\,,
\eeq
where $\pi^{(0)}_{\mu\nu}$ is the vacuum polarization term, which
is neglected, while the background dependent contribution is given by
\beq
\label{pim}
\pi^{(m)}_{\mu\nu} = -4e^2\int\frac{dp}{(2\pi)2E}
(f_e(E) + f_{\overline e}(E))
\left[\frac{L_{\mu\nu}}{k^2_\para + 2p_\para\cdot k_\para}
(k_\para \rightarrow -k_\para)\right]\,.
\eeq
In this formula,
\beq
\label{Lmunu}
L_{\mu\nu} = 2p_{\para\mu} p_{\para\nu} + p_{\para\mu} k_{\para\nu}
+ k_{\para\mu} p_{\para\nu} - g_{\para\mu\nu}p_\para\cdot k_\para \,,
\eeq
where $g_{\para\mu\nu}$ has been defined in \Eq{gparallel},
$f_{e,\overline e}$ denote the particle and antiparticle number density
distributions defined in \Eq{fe}, while $k^\mu_\para$ and $p^\mu_\para$
are parametrized as indicated in \Eqs{kparadef}{pparadef}, respectively. 
Furthermore, the integral in \Eq{pim} is to be interpreted in the sense of its
principal value part.

It is easily verified that, besides being symmetric and transverse to
$k^\mu_\para$, $\pi^{(m)}_{\mu\nu}$ satisfies
\beq
R^{\mu\lambda}\pi^{(m)}_{\lambda\nu} = 0\,.
\eeq
These properties imply that it is of the form given in \Eq{pieffform},
and the coefficient of $Q_{\mu\nu}$ can be found
by projecting \Eq{pim} with $Q_{\mu\nu}$. This procedure then yields
\beq
\label{piLm}
\pieff = 4e^2\frac{k^2_\para}{\kappa^2}B\,,
\eeq
with
\beqa
\label{B}
B & = & \int\frac{dp}{4\pi E}(f_{e}(E) + f_{\overline e}(E))\nonumber\\
&&\mbox{}\times
\left[\frac{2(p_\para\cdot u)^2 + 2(p_\para\cdot u)(k_\para\cdot u) -
p_\para\cdot k_\para}
{k^2_\para + 2p_\para\cdot k_\para}
+ (k_\para\rightarrow -k_\para)\right]\,.
\eeqa
Substituting \Eq{piLm} in \Eq{redispreleq}, we obtain
\beq
\label{dreqfinal}
\left(\log\left|\frac{k^2_\para}{\mu^2}\right|\right)^{-1}
- \frac{e^2}{\pi}\left(\frac{k^2_\para}{\kappa^2}\right) B = 0\,,
\eeq
which is the equation to be solved for $\omega(\kappa)$.

A useful formula for $B$, that holds when the photon momentum is such that
\beq
\label{lowmomentumlimit}
\omega, \kappa \ll E_e\,,
\eeq
where $E_e$ is a typical energy of an electron in the gas, is obtained
as follows. In this case \Eq{B} can be approximated by the form
\beqa
\label{Bksmall}
B(\omega,\kappa) & = & -\frac{1}{2}\int\frac{dp}{(2\pi)}
\left(\frac{v\kappa}{\omega - v\kappa}\right)
\frac{d}{d{E}}(f_e + f_{\overline e})\,,\nonumber\\
& = & -\frac{1}{2\omega}\int\frac{dp}{(2\pi)}
\frac{(v\kappa)^2}{\omega - v\kappa}
\frac{d}{d{E}}(f_e + f_{\overline e})\,,
\eeqa
where $v = p/E$ is the velocity of the particles in the background and
we have indicated explicitly the dependence of $B$
on the photon momentum variables. The form given in the second line
can be obtained from the first line by inserting in the integrand
the factor $(\omega - v\kappa + v\kappa)$.
\Eq{Bksmall} is obtained from \Eq{B} by expanding the integrand in
terms of $k_\para/E$ and retaining only the dominant terms when the limit
$k_\para/E$ is taken. If the gas in non-relativistic, \Eq{Bksmall}
holds for $\omega,\kappa \ll m_e$. For a relativistic gas, \Eq{Bksmall}
holds also for $\omega,\kappa > m_e$, subject to \Eq{lowmomentumlimit}.
Thus, \Eq{Bksmall} is a useful formula that can be employed to find the
dispersion relations from \Eq{dreqfinal} in many situations of interest.

\subsection{Dispersion relations}
\label{sec:disprel}

The integral in \Eq{Bksmall} cannot be reduced any further in general,
but it can be evaluated for specific cases of the distribution functions
or by making further approximations that depend on the kinematic
regime being considered and the conditions of the electron gas.


We consider for definiteness a completely degenerate electron gas.
A simple evaluation of \Eq{Bksmall} then yields
\beq
B = \left(\frac{v_F}{2\pi}\right)\frac{\kappa^2}{\omega^2 - v^2_F \kappa^2}\,,
\eeq
where $v_F$ is the Fermi velocity of the electrons. Substituting this in
\Eq{dreqfinal}, the dispersion relation, which we denote by
$\omega_\kappa$ is then obtained by solving
\beq
\label{dispreleqdeg}
\omega^2_\kappa - v^2_F\kappa^2 = \omega^2_p (\omega^2_\kappa - \kappa^2)\log
\left|\frac{\omega^2_\kappa - \kappa^2}{\mu^2}\right| \,,
\eeq
with
\beq
\label{omegapdegcase}
\omega^2_p = \frac{2\alpha v_F}{\pi} \,, 
\eeq
where we have introduced the fine structure constant $\alpha = e^2/4\pi$.
This equation has a solution for $\omega > \kappa$, but it is not
a physical one since it lies outside the range given in \Eq{lowmomentumlimit}
for which \Eq{Bksmall} is valid.

In order to consider the solution with $\omega < \kappa$,
we rewrite \Eq{dispreleqdeg} in the form
\beq
\label{dispreleqdeg2}
\omega^2_\kappa - v^2_F \kappa^2 = \omega^2_p(\kappa^2 - \omega^2_\kappa)
\log\left(\frac{\mu^2}{\kappa^2 - \omega^2_\kappa}\right)\,.
\eeq
The solution to this equation can be represented in parametric form as
\beqa
\label{parametricform}
(\omega_\kappa/\mu)^2 & = & \gamma^2_F v^2_F t +
\gamma^2_F \omega^2_p \log(1/t)\,, \nonumber\\
(\kappa/\mu)^2 & = & t + \gamma^2_F v^2_F t +
\gamma^2_F \omega^2_p \log(1/t) \,.
\eeqa
On the other hand, an explicit approximate solution of \Eq{dispreleqdeg2}
can be obtained as follows.
Since, according to \Eq{omegapdegcase}, $\omega^2_p$ is not larger than
about $10^{-2}$, we can consider an expansion in terms
of $\omega^2_p$. Thus, to the zeroth order,
\beq
\label{drzeroth}
\omega^{(0)}_{\kappa} = v_F \kappa\,,
\eeq
and substituting this term in the
right-hand side of \Eq{dispreleqdeg2} yields the solution
\beq
\label{drdeg}
\omega^{(1)}_{\kappa} = \left[v^2_F \kappa^2 +
\frac{\omega^2_p \kappa^2}{\gamma^2_F}
\log\left(\frac{\mu^2\gamma^2_F}{\kappa^2}\right)\right]^{1/2}\,,
\eeq
where
\beq
\gamma_F = (1 - v^2_F)^{-1/2} \,.
\eeq

\begin{figure}
\begin{center}
\includegraphics[bb=130 525 485 735,scale=0.70]{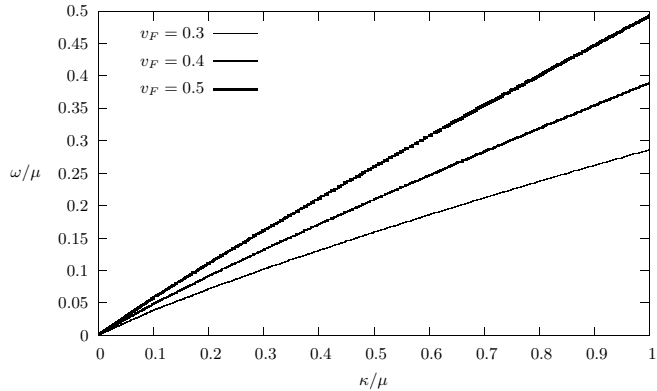}
\end{center}
\caption[]{Plot of the dispersion relation $\omega^{(1)}_\kappa$ 
defined in \Eq{drdeg}, for various values of $v_F$.
\label{fig:dr2}
}
\end{figure}

\begin{figure}
\begin{center}
\includegraphics[bb=130 525 485 735,scale=0.70]{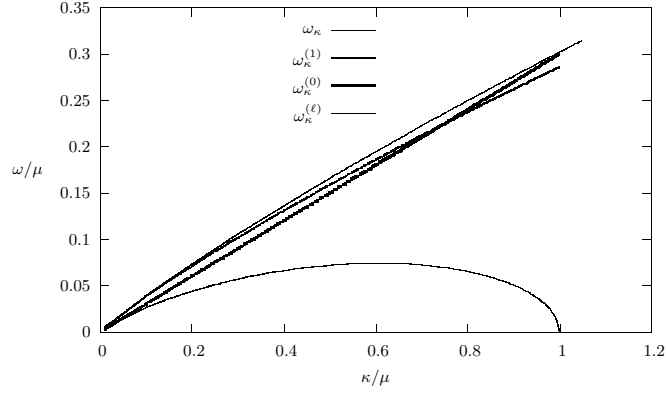}
\end{center}
\caption[]{Plot of the exact solution $\omega_\kappa$ 
of \Eq{dispreleqdeg2}, the analytic formula $\omega^{(1)}_\kappa$
given \Eq{drdeg}, the zeroth order term $\omega^{(0)}_\kappa$ defined in
\Eq{drzeroth} and the long-wavelength limit formula defined in
\Eq{drstaticapprox}, all for $v_F = 0.3$.
\label{fig:dr3}
}
\end{figure}

The function $\omega^{(1)}_\kappa$ is plotted in \Fig{fig:dr2}, for various
values of $v_F$. \Fig{fig:dr3} shows the plots
of $\omega^{(1)}_\kappa$ as well as the exact solution $\omega_\kappa$
of \Eq{dispreleqdeg2} together with the plot of the zeroth order
term $\omega^{(0)}_\kappa$, for $v_F = 0.3$.
For comparison, \Fig{fig:dr3} includes
the plot of the long-wavelength approximation formula
\beq
\label{drstaticapprox}
\omega^{(\ell)}_\kappa = 
\omega_p \kappa \left[\log\frac{\mu^2}{\kappa^2}\right]^{1/2} \,,
\eeq
which is quoted in the literature for the strictly 1D plasma.
\begin{figure}
\begin{center}
\includegraphics[bb=130 525 485 735,scale=0.70]{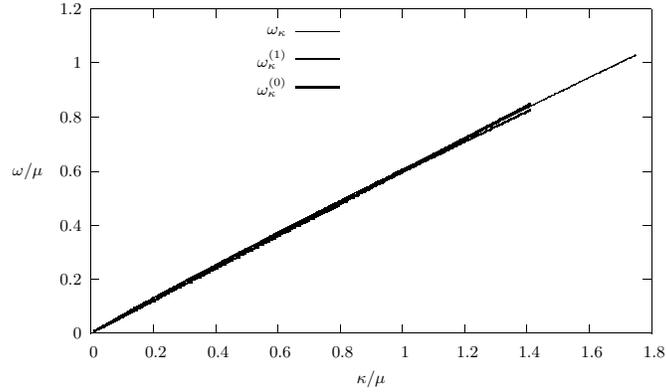}
\end{center}
\caption[]{
Plot of the exact solution $\omega_\kappa$ of
\Eq{dispreleqdeg2}, the analytic formula $\omega^{(1)}_\kappa$
and the zeroth order term $\omega^{(0)}_\kappa$, for $v_F = 0.6$.
\label{fig:dr4}
}
\end{figure}

Notice that \Eq{drstaticapprox} can be obtained
formally from \Eq{drdeg} when $v_F$ and $\kappa$ are small,
which evidences why that formula has a very restricted range of validity.
On the other hand, as can be seen from \Fig{fig:dr3}, the function
$\omega^{(1)}_\kappa$ approximates very well the exact solution
of \Eq{dispreleqdeg2}, represented by \Eq{parametricform}.
Their difference becomes smaller or even negligible for higher values of $v_F$,
as illustrated in \Fig{fig:dr4}.
The plots in \Fig{fig:dr2} exhibit the property that
the slope of the dispersion relation (the group velocity)
is almost constant for the entire range of $\kappa$,
which are some of the the general and unique characteristics
that have been observed in recent experiments in this
type of system\cite{kukushkin}.

The main result here is that a proper treatment of the strictly
1D plasma leads to a dispersion relation which, (1) is very different
from the long-wavelength formula of \Eq{drstaticapprox} that is quoted
in the literature for this system,
and (2) reproduces the expected characteristics that have been observed
in this kind of system. It is appealing that this result
has been obtained while insisting in considering the strictly 1D plasma.
This contrasts with previous treatments in the literature that
abandon the strictly 1D plasma and instead
use the fact that a semi realistic 1D plasma
has some finite width and therefore the calculations must
necessarily be numerical model calculations involving the details
of the electronic wave functions in the transverse directions in the wire.
%
%
\section{Conclusions}
\label{sec:conclusions}

The Thermal Field Theory methods have been used to study the propagation
of photons in the model of the strictly 1D plasma, that is,
a system in which the electrons that are free to move
but are confined to an infinitely thin tube,
or wire. An important step was to identify the appropriate
photon effective field and to determine the corresponding
free-field propagator. We performed the one-loop calculation of the photon
self-energy in that medium, and we considered the photon dispersion relations.

The dispersion relation was studied in some detail for the case
of a degenerate electron gas, and analytic formulas were obtained.
The dispersion relation has a leading term
$v_F \kappa$, where $v_F$ is the Fermi velocity of the electrons,
with logarithmic corrections proportional to $e^2 v^2_F/2\pi^2$.
This result is very different from the long-wavelength
formula that is usually quoted in the literature for this type of system.
In particular the formulas obtained here are valid for all the range
of values of $\kappa$ and they resemble the linear form
that have been observed in recent experiments this type of
system\cite{kukushkin}, and which are expected on the basis of
the numerical calculations that take into account
the electronic structure and finite-width of a real
quasi-1D system\cite{sarma}.

While we envisaged the simplest situation
of an ordinary gas of electrons, which are confined to a wire but are
otherwise free, the same method can be used to
to consider variations of the model in a systematic way, such as
the effects of external fields,
which have been studied by other means\cite{aleiner}.

The application of TFT that we have described here,
and to the layer in \Ref{nieves:layer}, can be useful in the
context of the recently proposed ``vanishing dimensions''\cite{stojkovic3}
and ``layered structure of space''\cite{stojkovic1} ideas.
The methods can be also useful
in astrophysical\cite{lyutikov}, plasma physics\cite{jiang}
and condensed matter\cite{nagao,uchida,kukushkin}
and other systems of current interest
in which a plasma is confined to a layer\cite{caldas} or a wire\cite{faccioli}.

\end{document}